\documentclass[12pt]{article}

\begin{document}
\title{Cutoff function in holographic RG flow}
\author{F. Ardalan$^*$\\	
School of Particles and Accelerators , \\ Institute for Research in Fundamental Sciences
 (IPM),\\
Tehran, P.O. Box 19395-5531, Iran\\
Physics Department, Sharif University of Technology,\\
Tehran,  P.O. Box 11365-11155,  Iran\\
$^*$  ardalan@ipm.ir}

\maketitle
 Using the precursor map in AdS/CFT, the renormalization group cutoff function is mapped to the dual theory. The resulting flow equations on the two sides of the duality are compared.

\newpage
\section{INTRODUCTION}
AdS/CFT duality between a conformal quantum field theory and gravity with an extra dimension, has occupied a large part of theoretical physics research in the past two decades [1-3] 

Moving away from the conformal fixed point in the field theory space, the duality is believed to persist once the gravity on the dual space is allowed to evolve along the extra dimension according to a specific recipe [4,5].

In this renormalization group (RG) holographic set up, the connection between the field theory energy cutoff scale and the extra dimension cutoff scale on the gravity side is not well understood [6]. Moreover,  the smooth cutoff function utilized in the field theory side as elaborated in [7],  has no obvious counterpart on the gravity side.

The present work addresses this question via the precursor map. The concept of the precursor originated from the observation that,  in AdS/CFT, an event in the bulk must be perceived in the boundary theory immediately and an operator be assigned to it there, thus the precursor [8].
Where the two sides of the duality are known, a map is constructed which assigns an operator in the conformal field theory to every field on the gravity side [9]:
\begin{equation}
\varphi(z,x)=\int d x' K(z,x\mid x') \mathcal{O}  (x')
\end{equation}
where $K$ is called the smearing function, a green function of the gravity theory, and $\mathcal{O}$ is an operator in the conformal field theory; $z$ is the extra dimension, the radial direction.

Using this map, it is possible, in principle, to relate the RG flow of the field theory side to the evolution along the extra dimension in the gravity side.
In particular,  it must be possible to relate the cutoff energy scale of the RG flow to the cutoff scale of the extra dimension in the gravity side and the smooth cutoff function in the RG equation [7] to a corresponding smooth cutoff function on the gravity side. However, it is not obvious how this will happen in an arbitrary RG scheme.

There is an RG scheme, though, in which the correspondence between the two cutoff functions is straightforward [10]. In this scheme, the momentum cutoff is directly imposed on the fields themselves via multiplication of the fields by a cutoff function in the momentum space. It is then only a matter of reading off the precursor map of the cutoff field in the gravity side and, conversely, given a cutoff function on the bulk side will lead to a cutoff operation on the field theory side.

It will be shown here that the outcome of the map is not a simple multiplication of the fields by a cutoff function, when it is so on the other side. Rather, it will turn out to be a type of convolution of the fields by the image of cutoff function. This is not unexpected as the map (1) involves a non-local smearing. The consequence of this outcome will be discussed in some detail, and the relation between the RG equation on the boundary theory with the evolution equations on the bulk theory side will be explored.This behavior has then the important consequence that there will not be a sharp cutoff on one side of the duality when the cutoff is sharp on the other side.
 
In Sec. II the holographic renormalization is briefly reviewed and the multiplicative renormalization scheme is introduced. Furthermore, the precursor map is explicated. Section III contains the main result of this investigation and is devoted to the study of the precursor map on the renormalization scheme, and its ramification on the flow equation and evolution equation, respectively, on the two sides of AdS/CFT duality. Section IV is devoted to speculations on the precursor map's relevance to the holographic renormalization group flow. 

After the initial submission of this work on the arXives, I was informed of the references [11], [12] in which the bulk reconstruction was carried out for the cutoff procedure of Refs. [4,5]. There, the map is constructed for on shell configurations of the bulk fields. In contrast, the cutoff function imposed on the precursor map in this article moves the configuration off shell, as it is to be utilized in the path integrals of the duality equivalence considered here. Therefore, one does not expect the map to reduce to that considered in [11,12] when the cutoff function is made sharp. This will be reflected in the absence of negative frequency Bessel functions in its expansion.

\section{HOLOGRAPHIC RENORMALIZATION GROUP}

In quantum field theory, the central quantity is the partition function

\begin{equation}
Z=\int D\varphi e^{-S (\varphi)}.
\end{equation}

In the AdS/CFT duality, the partition function of the boundary conformal field theory is related to the partition function of the bulk gravity theory through the transform:

\begin{equation}
Z_B \equiv \int D \varphi _{(z,x)} e^{-\scriptsize{s}(\varphi)}=\int D \Phi(x) e^{-\mathcal{S}(\Phi)+\int dx \varphi_0 (x) \mathcal{O} (x),}
\end{equation}
Where $Z_B$  is the bulk partition function, $\scriptsize{s}$ is the bulk action, $S$ the boundary action, $\varphi_0$ the boundary value of the bulk field, $\varphi_0 (x)=\varphi (z,x=0)$, and $ \mathcal{O}(x) $ is the corresponding boundary conformal field theory operator.

Correlation functions of the quantum field theory may be obtained by functional derivatives of the partition function in the presence of external sources $J$:

\begin{equation}<\varphi (x_1).... \varphi (x_n)>= \frac{\delta}{\delta J(x_1)} ... \frac{\delta}{\delta J(x_n)} Z_J,\end{equation}

\begin{equation}Z_J \equiv \int D \Phi  e^{-S(\Phi) + \int J \Phi dx}.\end{equation}

In calculating the path integrals for the correlations functions, one may first integrate over the high-momentum modes above a certain energy scale $\wedge$ and write the partition function $Z$ as\\

\begin{equation}Z= \int_{p< \wedge} D \Phi e^{-S_\wedge (\Phi)},\end{equation}
where $S_\wedge (\Phi)$ is now an effective action, which, by definition, would satisfy a differential equation, the RG flow equation.
This equation first written down by explicitly path integrating the higher-momentum modes, was given by Wegner and Houghton [13], which was subsequently generalized to the case of a smooth cutoff function by Wilson and Kogut[14] and later  Polchinski [7]:

\begin{equation}\wedge \frac{d}{\Phi \wedge} S_{int}= -\frac {1}{2}\int dp \wedge \frac{d}{d\wedge} K [ \frac{\delta S_{int}}{\delta \varphi (-p)}\frac{\delta S_{int}}{\delta \varphi (p)} +\frac{\delta^2 S_{int}}{\delta \varphi (-p)\delta \varphi (p)} ],\end{equation}
Where  \begin{equation}S_{\wedge} = -\frac{1}{2} \int \frac{\varphi(-p)\varphi(p)} {K}+ S_{int}.\end{equation}

The germ of the smoothing function $K$ is in the imprecise notion of "almost integration" of Wilson and Kogut [14]. The cutting off of the higher momentum modes in the path integral is accomplished in the Polchinski equation by the form of the cutoff function $K(p),$ which almost equals to 1 for momenta less than $\wedge$ and rapidly decays to zero for momenta larger than $\wedge$. Clearly the resulting correlation functions will depend on the form of $K$. But a judicious choice of the function renders the dependence minimal. An optimum choice is one which is nearly a step function.  In the rest of this article, such a choice of cutoff function is to be understood.

The   imposition of the cutoff function on the kinetic term of the action makes it cumbersome for certain purposes as for the precursor   map.
There is an alternative method of imposition of cutoff on the momentum modes which will be natural for the precursor map considerations.
In this scheme, the momentum space fields are directly multiplied by a cutoff function,  so that the  higher-momentum modes are "almost deemphasized" in the fields, to begin with [10],
\begin{equation}\varphi (p) \longrightarrow h(p) \varphi (p),\end{equation}
where $h(p)$ is a function which is nearly 1 for momenta less that a scale $\wedge$, and rapidly vanishing for momenta larger than $\wedge$.

The resulting RG equation, following the usual procedure, turns out to be similar to Polchinski equation, and is  \footnote{ The equation for the total action in this regularization scheme turns out to be identical to that of the interaction part of action in the Polchinski-Wilson scheme   (7); detail may be found in Ref. [15].} 

\begin{equation}\wedge \frac{d}{d \wedge} S_{\wedge}= -\frac{1}{2} \int dp \wedge \frac{d}{d\wedge} h [ \frac{\delta S_{\wedge}}{\delta \varphi (p)}\frac{\delta S_{\wedge}}{\delta \varphi (-p)} +\frac{\delta^2 S_{\wedge}}{\delta \varphi (-p)\delta \varphi (p)} ]\end{equation}

In   AdS/CFT, the question of the relation between the two sides of the duality when the boundary field theory undergoes renormalization group flow was addressed almost a decade ago, [4,5]. The conjecture is that, when the boundary field theory  is cutoff at a scale $\wedge$, the corresponding dual gravity theory is cutoff along the extra dimension with the scale $z_0$, where

\begin{equation}\wedge \sim \frac{1}{z_0}.\end{equation}

Explicity, the partition function of the gravity side  is related to the generating functional on the field theory side as follows: 

\begin{equation}Z_{B, z_0} \equiv \int_{z>z_0} D \varphi (z,x) e ^{-\scriptsize{s}(\varphi)}= \int D \Phi(x) e^{-{S}_{\wedge}(\Phi) + \int \varphi_0 (x) \mathcal{O}(x)}\end{equation}
Here ${S}_\wedge (\bar \Phi)$ is the boundary field theory cutoff at $\wedge$,  $\scriptsize{s} (\varphi)$ is the bulk action, and $\varphi_0 (x)=\varphi (z=z_0, x)$ is the bulk field  at $z=z_0$.

The precise dependence of $z_0$ on $\wedge$ is not known [6]. Also, it is not obvious how a smooth cutoff function in the boundary field theory RG is projected on a similar mechanism on the bulk side. Here the precursor map may come to the rescue. If one knows how objects on the two sides of duality correspond to each other, then it may be possible to find what corresponds in the bulk theory to the RG flow of the boundary field theory, and, hopefully, find the precise relation between the aforementioned cutoff scale $\wedge$ of the boundary field theory and the cutoff scale of the extra dimension in the gravity bulk theory. From the very early beginnings of AdS/CFT duality, it was understood that the information in the bulk should be somehow encoded in the boundary theory and, in particular, events in the bulk should be represented in the boundary conformal field theory [8]. Thus, a field operator in the bulk should be given by an operator in the boundary, called the precursor. A generic precursor map   (1)  was constructed by Hamilton, Kabat, Lifschytz, and Lowe (HKLL) giving a bulk field $\varphi(x,t)$ as an operator in the boundary theory[9].

In the particular  case of the dual pair of the $O(N)$ vector model in three dimensions, and the higher  spin theory in the four-dimensional bulk, the smearing function $K$ turns out to be a set of delta functions in the momentum representation [16]. The exact expression will be written down in the next section.

In general, K is in term of the mode functions of the bulk space, for pure AdS, being Bessel functions [9].
\section{PRECURSORS AND CUTOFFS}
 
 In this section, the precursors are used for the multiplicative regularization scheme to relate the  cutoff functions of the two sides of the  AdS/CFT duality and relate the two evolution equations.
 
 Generally for a scalar theory the HKLL [9] reconstruction gives the precursors in the form   (1). Therefore, the  multiplicative regularization of the boundary fields
 \begin{equation}
 \tilde{\mathcal{O}}(p)\longrightarrow  h(p)\tilde{\mathcal{O}(p)}, \end{equation}
 where $h(p)$ is a generic smooth cutoff function on a generic boundary momentum space operator, is mapped via this precursor map to a corresponding "cutoff" field $\varphi_h (x)$ in the bulk. This cutoff field will certainly not be a sharp cutoff field in the bulk even when $h(p)$ is a sharp cutoff. To see its behaviour, the concrete form of the smearing function should be inserted in the expression (1):
 
 \begin{equation}
 \tilde{\varphi}_h (p)= \int dp' \tilde{K} (z, p|p') h(p') \tilde \mathcal{O} (p').
 \end{equation}
 
 Conversely, following the holographic renormalization conjecture [4,5], the precursor may allow   finding the behaviour of the cutoff field ${\tilde \mathcal{O}_\rho} (p)$ in the boundary theory:
 
\begin{equation}
\rho(z) \tilde{\varphi}(p)= \int dp' \tilde{K} (z,p |p'){\tilde \mathcal{O}_\rho} (p),
\end{equation}
 where $\rho(z)$ is the cutoff function on the z direction of the bulk theory, and it is a sharp cutoff in Refs. [4,5].
 
 It is, in general,  not   straightforward to find the induced cutoff functions in either case. However, in the particular case of $O(N)$/higher-spin AdS/CFT duality, it is possible to find these cutoff functions and, therefore, see how a certain cutoff on one side of the duality is affected on the other side and, in particular, find the behaviour of the cutoff when on the other side  the cutoff is sharp as in Ref. [4]. This latter behaviour is further studied from now on.
 
 In the $O(N)$ /higher-spin duality the operator on the CFT side is the bi-local field $\mathcal O(x_1, x_2)$:
 
 \begin{equation}
 \mathcal{O} (x,y) \equiv \Phi_i (x) \Phi_i (y)
 \end{equation}
 where $\Phi_i(x)$ are the boundary $\mathcal{O} (N)$ fundamental fields.
The precursor map between the bulk higher-spin fields $\varphi(z,x)$ and the boundary bi-local  fields $\mathcal O(x_1, x_2)$  are best written down  in momentum space and are [16], 

\begin{equation}
\tilde{\varphi}_s(z,x)=\int d^4 p ~~~e^{i(p.x+p^z z)} \int d^2 P_1 d^2 P_2 \delta (p^+_1 + p^+_2 -p^+) . \delta (p_1 +p_2-p) . \delta (p_1 \sqrt {\frac{p^+_2}{p^+_1}} - p_2 \dot{\sqrt{\frac{p^+_1}{p^+_2}}} - p^z ) .\end{equation}

 $(\frac{1}{p^+_1} 
+ \frac{1}{p^+_2} ) (p^+_1 + p^+_2)^s P_s^{\frac {-1}{2}, \frac{-1}{2}} 
(\frac {p_2^+ - p_1^+}{p_2^+ + p_1^+}) . \tilde{\mathcal{O}} (p_1, p_1^+ ; p_2, p_2^+),$\\
where $\varphi_s$ is the bulk highe-spin field of spin $s$ and $z$ the radial direction; $P_i= (p_i, p^+_i, p^-_i)$, $i=1,2$, are the momentum light cone variables of the $O(N)$ boundary fields $\phi(p_i, p^+_i)$, and $P_s^{\frac {-1}{2}, \frac {-1}{2}}$ are the Jacobi polynomials.
This relation can be put in the HKLL form [16]:

\begin{equation}
\tilde{ \varphi_s (z,x)}=\int_{p^2<0} d^3 p e^{ip.x} J_{-\frac{1}{2}}  (z\sqrt{-p^2}).\sqrt{{\frac{\pi z}{2}}\sqrt {-p^2}}~~\frac{s!}{\Gamma (s+\frac{1}{2})}~~\frac{1}{(p^+)^s}\tilde{\mathcal O}_s (p),
\end{equation} 
where now $ \tilde \mathcal{O}_s $ are the CFT boundary operators of HKLL and are 

\begin{equation}
\tilde{\mathcal O}_s (p) = \int d^2 P_1 d^2P_2 (\frac {1} {p^+_1} + \frac{1} {p^+_2} ) \delta (p^+_1 +p^+_2 - p^+) \delta(p_1+p_2-p) . (p^+_1 + p^+_2)^s P_s^{-\frac{1}{2}, -\frac{1}{2}}~~ (\frac{p_2^+ - p_1 ^+}{p_2^+ + p_1 ^+}) ~~\tilde{\mathcal{O}} (P_1, P_2)
\end{equation}
in terms of the bilocal fields $\mathcal{O}(x_1, x_2)$.\\
It is now possible to trace the effect of a cutoff on the bulk radial direction of the fields $\varphi$, on the boundary operator $\mathcal{O}_s$, when

\begin{equation}
 { \varphi_s(z,x)}\to   \varphi_s^\rho ~~ (z,x)\equiv\rho (z) \varphi_s (z,x),  
\end{equation}
 where $\rho (z)$ is some smooth cutoff function in the z direction.
In Eq. (17), changing the $p_1$ and $p_2$ variables in the integral to $p_1+p_2,~ p_1^+ + p_2^+,$ ${\frac {p^+_2 - p^+_1}{p^+_2 + p^+_1}}$, and

\begin{equation}
p_1 \sqrt{\frac{p^+_2}{p^+_1}} - p_2 \sqrt{\frac{p^+_1}{p^+_2}} \equiv Q, 
\end{equation}
and taking into account its Jacobian, it is simple to see that regularization of $\varphi_s$ with  the multiplication by $\rho (z)$, [Eq. (20)],  induces a convolution transformation of $\tilde{ \mathcal{O}}$ by $\tilde{\rho} (p^+)$ along the z direction:

\begin{equation}
\tilde{ \mathcal{O}}(... , Q) \to \tilde{ \mathcal{O}}_{\rho} (... , Q)\equiv (\tilde{\rho} o  \tilde{ \mathcal{O}})(... , Q).
\end{equation}
Then using the relation

\begin{equation}
p^2 + (p^z)^2=0, 
\end{equation}
it is immediate that the spin s-primary operator of the boundary theory, $\mathcal{O}_s$, is similarly regularized:

\begin{equation}
\tilde{\mathcal{O}}_s(p)\rightarrow ({\tilde \rho} o  \tilde \mathcal{O}_s)(p).
\end{equation}

This is not exactly the cutoff procedure in exact renormalization group flow treatment in field theories. However, it may become of a form more familiar in RG; by taking the dual Fourier transform of   Eq.(24)  one ends up with

\begin{equation}
\tilde {\tilde{\rho}} (r) \tilde{\tilde{\mathcal{O}}}(r),
\end{equation}
where r is the "dual Fourier" coordinate of p. This is reminiscent of the internal coordinate of the bilocal system considered in [6].

The case of a sharp cut off in the bulk, 

\begin{equation}
\rho(z)=1 ~~~~for~~~ z_0< z < z_{\infty}, ~~~ zero ~~~ otherwise,
\end{equation}
is of particular interest. Here $z_0$ is the UV cutoff point and $z_{\infty}$ is an IR cutoff. Then

 \begin{equation}
\tilde {\rho} (p) \sim e^{\frac{i}{2}(z_{\infty} + z_0)p} ~~~~\frac{sin p \Delta z}{p},
 \end{equation}
  where $\Delta z= ~~ z_{\infty} - z_0$.\\
  As $z_0 \to z_\infty$, the width of $\tilde{\rho}(p)$ increases and the operator 
  
  \begin{equation}
 (\tilde{ \rho}o\tilde{\mathcal{O}})(p) = \int dp' \tilde{\rho} (p-p') \tilde{\mathcal O}^{\rho} (p'), 
  \end{equation}
  at a fixed  p, becomes 'larger',  and $\tilde{\mathcal{O}}$ is partially integrated out. In the limit  $z_0 \to z_{\infty}$, $\rho \to \delta$ and
  
   $$(\tilde{ \rho}o\tilde{\mathcal{O}})(p) \to \int \tilde{ \mathcal O}(p')dp';$$
  while for $z_0 \to 0$, $\rho \to 1$, and $\tilde{\rho} \to \delta$;
  then $( \tilde{\rho}o \tilde{\mathcal{O}})(p) \to \tilde{\mathcal{O}}(p)$   and no modes are integrated out.
  
In the above construction, the precursor map was entirely derived from symmetry considerations [16], which uses a map between the canonical coordinates of the two dual theories. In other cases, the map is derived from the dynamical equation of the bulk theory as in the case below. However, in the above case, the map derived from dynamics is identical to that from symmetry arguments, as it should be [Eq. (18)].

Another example is the $AdS_3/CFT_2$ duality of a scalar field in the AdS background dual to a boundary conformal field, where the bulk theory
  
   \begin{equation}
S=\int dx^2dz~(\partial_\mu~ \varphi ~\partial_\mu ~\varphi ~+ m^2 \varphi^2 )
 \end{equation}
 is dual to a boundary  CFT of dimension $\Delta$ given by [3]
  
  $$\Delta(\Delta-2)= m^2.$$
  The bulk field is then related  to the boundary operator $\mathcal{O}$ as [9,17]
  
    \begin{equation}
\varphi (z,x)=cz  \int_{p^2<0} dp^2 e^{ip.x}~~ \frac{1}{p\nu} J_{\gamma} (pz) \tilde{\mathcal{O}} (p), 
  \end{equation}
  where $\nu=\Delta -1$. Here $p= \sqrt {p_\mu  p^{\mu} },$ and c is some numerical constant.
  
  This precursor formula makes it very easy to study the cutoff functions on the two sides of the duality. Using the orthognality relation of Bessel functions,
  
    \begin{equation}
  \int_0 ^{\infty} zp J_\nu (pz) J_\nu (p'z) dz= \delta (p-p'),
  \end{equation}
  the behavior of the cutoff CFT operator $\mathcal{O}$ is obtained:
 
\begin{equation}
\frac{1}{p\nu} \tilde{\mathcal{O}}_{\rho}(p)=\int_0 ^{\infty} dp' I_{\rho} (p, p')\frac{1}{p^{'\nu}}  \tilde{\mathcal O}(p'),
 \end{equation} 
 where   
 
    \begin{equation}
I_ \rho (p,p')=\int dz pz J_\nu  (pz) J_{\nu} (p' z) \rho(z).
  \end{equation}  
  
   Again, this cutoff operator (32) is not a simple product of the operator $\mathcal{O}$ by a cutoff function;
   rather it is a smearing operation on it. But, the integral operator $I(p, p')$ has the general properties expected from a genuine cutoff procedure. Indeed, at large momentum, $\tilde{\mathcal{O}_{\rho}}$ is damped, and it is not damped at smaller momenta.  To be precise, consider a cutoff function $\rho(z)$ which  decreases rapidly at $z<z_0$; then the properties of the Bessel functions $J_\nu$  imply that in Eq. (35),  for momenta p or p' larger than $\frac{1}{z_0}$, the function $I_{\rho} (p, p')$ is small, compared to when momenta have  finite values.
   
   The result (32) leads to an explicit formula for the boundary cutoff operator $\mathcal{O}_\rho$, in terms of Bessel function, when $\rho(z)$ is a sharp cutoff function on the bulk fields as in the original [4,5] holographic renormalization conjecture. When $\rho(z)=0$ for $z<0$ and 1 otherwise, Eq.  (33) becomes

 $$I(p,p')= \int_{z_0}^{\infty}pz~dz J_\nu (pz)
 J_\nu (p'z)$$ 
$$~~~~~~~~~~~~~~~~~~~~ = \delta(p,p')-\int_0 ^{z_0}pz dz J_0 (pz) J_\nu (p'z)$$
 \begin{equation}
  ~~~~~~~~~~~~~~~~~~~~~~~~ =\delta(p-p')-\frac{pz_0}{p^2-p'^2}~~[ p' J_\nu (pz_0)J'_\nu (p'z_0) - pJ'_\nu (pz_0)J_\nu (p'z_0)].
   \end{equation}
 \\
One can see from this relation   that there is a soft cutoff at momenta of the order of  $>\frac{1}{z_0}.$
 
   There remains to compare the evolution equations on the two sides of the duality. The generalization of AdS/CFT duality equivalence conjecture [ Eq. (12)], is

\begin{equation}
{Z_\rho} (\varphi_0)= e^{-W_\rho (\varphi_0)},
\end{equation}
  where $Z_\rho$ is the bulk partition function, 
  
  \begin{equation}
  Z_\rho(\varphi_0)= \int D\varphi (z,x) e^{-\scriptsize{s} (\rho \varphi)},
  \end{equation} 
and $e^{-W_\rho}$  is the generating  functional of the boundary field theory:

\begin{equation}
e^{-W_\rho (\varphi_0)}= \int D\Phi(x) e^{-S(\Phi_\rho)+ \int\varphi_{0\rho}\Phi_\rho}.
\end{equation}
   Here, and $\varphi_{o \rho}$ is the corresponding cuttoff boundary operator, $\varphi_{0,\rho}= I_\rho \varphi_0$, and $\varphi_0$ is the bulk field at the inflection point of the  cutoff function $\rho(z)$:
   
   \begin{equation}
  \varphi_0 (x)=\varphi(z=z_0,x),
   \end{equation}
   with $z_0$  the inflection point of $\rho.$\\
   Also, $\Phi_\rho (x)$ is the cutoff boundary field induced by the precursor map,   Eq. (32). As the cutoff function is changed, the two sides of the equivalence equation (35) will change according to the two evolution equations, as follows.
   
   On the bulk side the equation was derived in Refs.[4,18] and is the Schrodinger equation
   
   \begin{equation}
\dot{Z_\rho} (\varphi_0)= \mathcal{H} (\varphi_0, \frac{\partial}{\partial \varphi_0})Z_\rho (\varphi_0).
   \end{equation}
   Here $\mathcal{H}$ is the "Hamiltonian" derived from the bulk action $\scriptsize{S}$, with the radial direction replacing time.
   Strictly speaking, this equation is valid in the case of  $z_0=0$ or when $\rho(z)$ is sharp. But, as $\rho$ is always considered to be nearly sharp, this equation is almost valid in the spirit of "almost path integration" of Wilson, alluded to in the introduction.
   
   The evolution equation on the field theory side, of the generating functional, is related to the RG equation in the theory and can be obtained by similar techniques [7,13,19,20].\\
   Differentiate Eq.(37) with respect to $z_0$:

$$(e^{-W_\rho(\varphi_0)}\dot{)}=\int D\Phi [-\frac{\partial}{\partial z_0}  S ( \frac{\partial}{\partial \varphi_0 \rho})-\dot{I_\rho}\Phi  S' (\Phi_\rho) +\varphi_{0} ( I^2 _\rho \dot{)}~\Phi ]e^{-S(\Phi_\rho) + \varphi_{0\rho}\Phi_\rho}$$

\begin{equation}
~~~~~=\int dp [-\frac{\partial}{\partial z_0}  S ( \frac{\partial}{\partial \varphi_0 \rho})-\dot{I_\rho} \frac{\partial}{\partial \varphi_0}  S' ( \frac{\partial}{\partial \varphi_0 \rho})+ +2\varphi_{o}  \dot{I_\rho}\frac{\partial}{\partial \varphi_{0\rho}}]e^{-W_\rho (\varphi_0)},
   \end{equation}
   where the prime on S indicates differentiation with respect to the field. As in  Eq. (32),  the cutoff boundary field $\Phi_\rho$ is defined by 
   
   \begin{equation}
 \Phi_\rho(p)=\int dp' I_\rho (p,p')\Phi(p'),
   \end{equation}
or, generally,

   \begin{equation}
\Phi_\rho  \equiv I_\rho \Phi,
   \end{equation} 
   where $I_\rho$ is the cutoff linear operator induced on the boundary fields by the bulk cutoff function $\rho$. But,  $ \frac{\partial}{\partial z_0}  S  $ is obtained from the generalization of Polchinski-Wilson   [20]
   \begin{equation}
  \frac{\partial}{\partial z_0}  S  = -\frac{1}{2}\int dp  [\frac{\partial}{\partial \varphi}  S\dot{I_\rho} \frac{\partial}{\partial \varphi } S + \frac{\partial}{\partial \varphi}  \dot{I_\rho}\frac{\partial} {\partial \varphi }S]. \end{equation}
   Throughout, the relation $f_\rho.g_\rho \equiv \int dp(I_\rho f)(p)(I_\rho g)(-p)=\int dp~f (I_\rho)^2 g$ is understood. And the distinction between the CFT operator $\tilde {\mathcal{O}}$ and its modification $\frac{1}{p^\nu} \tilde{ \mathcal O}(p)$ is overlooked.
   
   The duality conjecture equation (35) relates the dynamics of the two sides of the duality from equating  (39) and (40):
   
   \begin{equation}
\mathcal{H} (\varphi_0, \frac{\partial}{\partial \varphi_0})=  -\frac{\partial}{\partial z_0}  S ( \frac{\partial}{\partial \varphi_0 \rho}) -\dot{ I_\rho} \frac{\partial}{\partial \varphi_{0}} S' 
(\frac{\partial}{\partial \varphi_{0\rho}})+2\varphi_{o}  \dot{I_\rho}\frac{\partial}{\partial \varphi_{0\rho}}.
   \end{equation}
   
   From this equation, one may derive the Hamiltonian of the bulk theory from the knowledge of the action of the boundary theory. There is an important caveat here; one assumes the form of the background in the bulk and, thus, the form of the integral operator $I_\rho$ which is in terms of the mode functions in the bulk background. Of coure, a central assumption is the validity of the precursor map. Moreover, there is a great deal of freedom in the choice and support of the precursor map which complicates application of the relation (44), [21].
   
   \section{CONCLUSION}
   In this work, a precursor map between the bulk fields and boundary operators was used to study the relation between the cutoffs on the bulk and boundary theories and to relate the evolution equation in the radial direction of the bulk theory to the RG flow equation of the boundary theory.
   To do this, a smooth cutoff  function on the the radial direction was imposed and the consequent cutoff operation on the boundary theory obtained. It was found that, generally, the operation of cutting of higher momenta in the boundary field theory is some convolution and that a sharp cutoff in the bulk  does not lead to a sharp cutoff on the field theory momenta.
   
   It is interesting to find out,  in the reverse direction, what the cutoff functions on the field theory in the boundary lead to   in the bulk theory.
   
   The introduction of smooth cutoffs on both sides of the duality allows one to find the evolution equations on the respective theories. On the bulk side, taking the bulk field's value at the inflection point of the cutoff function as the boundary value $\varphi_0$, it is straightforward,  from previous studies, to arrive at the Schrodinger equation for the evolution of the bulk partition function as the inflection point $z_0$ varies.
   
   But on the field theory side, the evolution of the generating functional gives an equation (40) which is not, strictly speaking, the usual RG equation  (10)  of the field theory. Equating the differential operators on the two sides of the duality acting on the single functional of the duality equivalence [Eq. (35)] leads to a relation  [Eq. (44)] between the Hamiltonian in the bulk and the action functional  of the boundary. As there is no unique  RG equation,  equation 44 is not a unique relation between  the dynamics of the two sides of the duality. Yet, given the dynamics on one side, a dynamics for the other can be derived from this equation.
   
   One should not forget that the usual equivalence, referred to as Hamilton-Jacobi equation versus RG equation, is between the nonrelativistic limit of the bulk Schrodinger equation and the Calan-Szymansik equation of the boundary.
   
      It is of interest to study this equation in concrete examples of AdS/CFT duality.

In a separate development, it has been conjectured [22] that, in the case $AdS_3/CFT_2,$  there is a duality between the cutoff bulk theory and the CFT boundary theory modified by a $T\bar T$
    term, where T is the energy momentum tensor component $T_{zz}$,  in the complex coordinate system.  Later [23], it was shown that, both in two dimensions and also in higher dimensions with the appropriate generalization of the  $T\bar T$ term, the CFT modification can be derived from the Hamilton-Jacobi equation of the bulk theory, in a similar spirit as that of this article. 

It is suprising that such a simple modification is all that remains from a large collection of possible irrelevant operators in moving away from the CFT fixed point.

This point will be addresses in future work.

   \section{ACKNOWLEDGEMENT}
   
   The author would like to thank Ali Naseh, Hessammadin Arfaei and Amin Faraji for discussions and reading of the manuscript and for collaboration in the early stages of this work with Amir Esmaeil Mosaffa and Shima Asnafi.

\end{document}